\documentclass[conference, 10pt]{IEEEtran}

\usepackage[T1]{fontenc}
\usepackage[utf8]{inputenc}
\usepackage[english]{babel}
\usepackage{balance}
\usepackage{amsmath, mathtools}
\usepackage{amsfonts,amsthm,bm}
\usepackage{amssymb}
\usepackage{graphicx}
\usepackage{standalone}
\usepackage{mathtools}
\usepackage{color, colortbl}
\usepackage{soul}
\usepackage[acronym,shortcuts]{glossaries}
\usepackage{tikz}
\usepackage{arydshln}

\newacronym{ar}{AR}{achievable rate}
\newacronym{awgn}{AWGN}{additive white Gaussian noise}
\newacronym{bd}{BD}{block diagonalization}
\newacronym{df}{DF}{decode-and-forward}
\newacronym{gnb}{gNB}{new-generation node base}
\newacronym{ga}{GA}{genetic algorithm}
\newacronym{los}{LoS}{line-of-sight}
\newacronym{mimo}{MIMO}{multiple-input multiple-output}
\newacronym{miso}{MISO}{multiple-input single-output}
\newacronym{siso}{SISO}{single-input single-output}
\newacronym{mu-mimo}{MU-MIMO}{multi-user multiple-input multiple-output}
\newacronym{su-mimo}{SU-MIMO}{single-user multiple-input multiple-output}
\newacronym{irs}{IRS}{intelligent reflecting surface}
\newacronym{snr}{SNR}{signal-to-noise ratio}
\newacronym{svd}{SVD}{singular value decomposition}
\newacronym{ue}{UE}{user equipment}
\newacronym{af}{AF}{amplify-and-forward}
\newacronym{hd}{HD}{half-duplex}
\newacronym{fd}{FD}{full-duplex}
\newacronym{6g}{6G}{sixth generation}
\newacronym{mmwave}{mmWave}{millimetre-wave}
\newacronym{2d}{2D}{two-dimensional}
\newacronym{rf}{RF}{radio-frequency}
\newacronym{csi}{CSI}{channel state information}
\newacronym{nlos}{NLOS}{non-line-of-sight}
\newacronym{aoa}{AoA}{angle of arrival}
\newacronym{aod}{AoD}{angle of departure}
\newacronym{ula}{ULA}{uniform linear array}
\newacronym{upa}{UPA}{uniform planar array}

\DeclareMathOperator*{\argmax}{\arg\!\max}
\DeclareMathOperator*{\argmin}{\arg\!\min}

\DeclareMathOperator*{\diag}{diag}

\title{Maximum-Rate Optimization of Hybrid Intelligent Reflective Surface and Relay Systems}
\author{Alberto Rech, Federico Moretto, and Stefano Tomasin \\ \small Department of Information Engineering, University of Padova, Italy. 
\\Emails: rechalbert@dei.unipd.it, federico.moretto.1@unipd.it, tomasin@dei.unipd.it}

\begin{document}

\bstctlcite{IEEEexample:BSTcontrol}

\maketitle

\begin{abstract}
We consider a wireless communication system,  where a transmitting source is assisted by both a reconfigurable \ac{irs} and a \acl{df} \acl{hd} relay ({\em hybrid \ac{irs}-relay scheme}) to communicate with a destination receiver. All devices are equipped with multiple antennas, and transmissions occur in two stages. In stage 1, the source splits the transmit message into two sub-messages, transmitted to the destination and the relay, respectively, using \acl{bd} to avoid interference. Both transmissions will benefit from the \ac{irs}. In stage 2, the relay re-encodes the received sub-message and forwards it (still through the \ac{irs})  to the destination. We optimize power allocations, beamformers, and  configurations of the \ac{irs} in both stages, in order to maximize the achievable rate at the destination. We compare the proposed hybrid approach with other schemes (with/without relay and \ac{irs}), and confirm that high data rate is achieved for the hybrid scheme in case of optimal \ac{irs} configurations.
\end{abstract}


\glsresetall

\IEEEpeerreviewmaketitle

\section{Introduction}
 
A reconfigurable \ac{irs} is a programmable metasurface that can alter the phase and amplitude of an impinging signal by dynamically adjusting the reflection coefficients of its elements. Recently, \acp{irs} have drawn enormous research interest as a promising technology for the \ac{6g} of cellular networks \cite{9387701}, due to their ability of controlling the wireless propagation environment. Before the advent of \acp{irs},  relays have been studied and used in cellular networks to increase coverage and improve the received signal quality. Among various solutions, \ac{df} relays are \ac{hd} devices that alternate two stages, one wherein they receive a message from the source, and a second wherein they re-encode the message and transmit it to the destination. 

The alternative use of \acp{irs} and relays has been widely investigated.  
In \cite{Bjornson19},  \acp{irs} and single-antenna \ac{df} relays are compared in terms of power consumption, whereas in \cite{Huang19}  the energy efficiency of systems using \acp{irs} is compared to a system  with multi-antenna \ac{af} relays. A  comparison between \acp{irs} and \ac{df} \ac{hd}/\ac{fd} relays is presented in \cite{DiRenzo20}, proving that sufficiently large \acp{irs} yield higher spectral and energy efficiency than relay-aided systems. 
Nevertheless, due to the expensive deployment of \acp{irs}, {\em hybrid \ac{irs}-relay systems}, wherein both devices are jointly adopted, will be a cost-effective solution for the near future of smart electromagnetic environments.
In \cite{Abdullah20}, the combination of a \ac{hd} \ac{df} relay and an  \ac{irs} is investigated and tight upper bounds for the \ac{ar} are derived. A hybrid system with a \ac{fd} \ac{df} relay is studied in \cite{Abdullah21}, showing that the performance  further improves, as long as the relay self-interference is low. However, both works consider source and destination equipped with a single antenna each. In \cite{9335947}, a system wherein an \ac{irs} assists both a relay and a destination (and the source has no direct link with either the relay and the destination) is considered, with source, relay, and destination again having all one antenna each. For a system with multiple relays, still in the presence of an \ac{irs}, the selection of one relay to assist communication between a source and a destination is solved by machine-learning  in \cite{9344820}.    
 
In this paper, we  consider a hybrid \ac{irs}-relay \ac{mimo}  system, which generalizes the systems considered in \cite{Abdullah20,Abdullah21}, and \cite{9335947}, as we now assume that all devices are equipped with multiple antennas. Moreover, contrary to  \cite{9335947}, we also consider the link between the source and the relay. The relay is \ac{hd} and operates in the \ac{df} mode. We propose a transmission protocol operating in two stages. In stage 1, the source splits the transmit message into two sub-messages, transmitted to the destination and the relay, respectively, using \acl{bd} to avoid interference. Both transmissions will benefit from the \ac{irs}. In stage 2, the relay re-encodes the received sub-message and forwards it (still through the \ac{irs})  to the destination. We optimize power allocations,  beamformers, and  configurations of the \ac{irs} in both stages,   to maximize the \ac{ar} at the destination. In particular, we split the  \ac{ar} optimization problem into two sub-problems, one for each stage,   then coupled by the choice of the \ac{irs} configuration and the power split between the signal for the relay and the destination in  stage 1. Lastly, we compare the proposed hybrid approach with other schemes (with/without relay and \ac{irs}), and confirm that a high data rate is achieved for the hybrid scheme in case of optimal \ac{irs} configuration.

The rest of this paper is organized as follows. Section II describes transmission characteristics and the  two-stage protocol. In Section~III we  formalize the maximum-rate optimization problem and describe the alternating optimization solution. In Section~IV we discuss numerical results before the conclusions are taken in Section V.

  {\it Notation:} Scalars are denoted by italic letters, vectors and matrices  by boldface lowercase and uppercase letters, respectively, and sets are denoted by calligraphic uppercase letters. $\diag(\bm{a})$ indicates a square diagonal matrix with the elements of $\bm{a}$ on the principal diagonal. $\bm{A}^H$ denotes the conjugate transpose of matrix $\bm{A}$. $\mathbb{E}[\cdot]$ denotes the statistical expectation. $\bm{I}_x$ is the identity matrix of size $x$.
 
\section{System Model}

\begin{figure}
 \centering
 \includestandalone[mode=buildnew, width=\columnwidth]{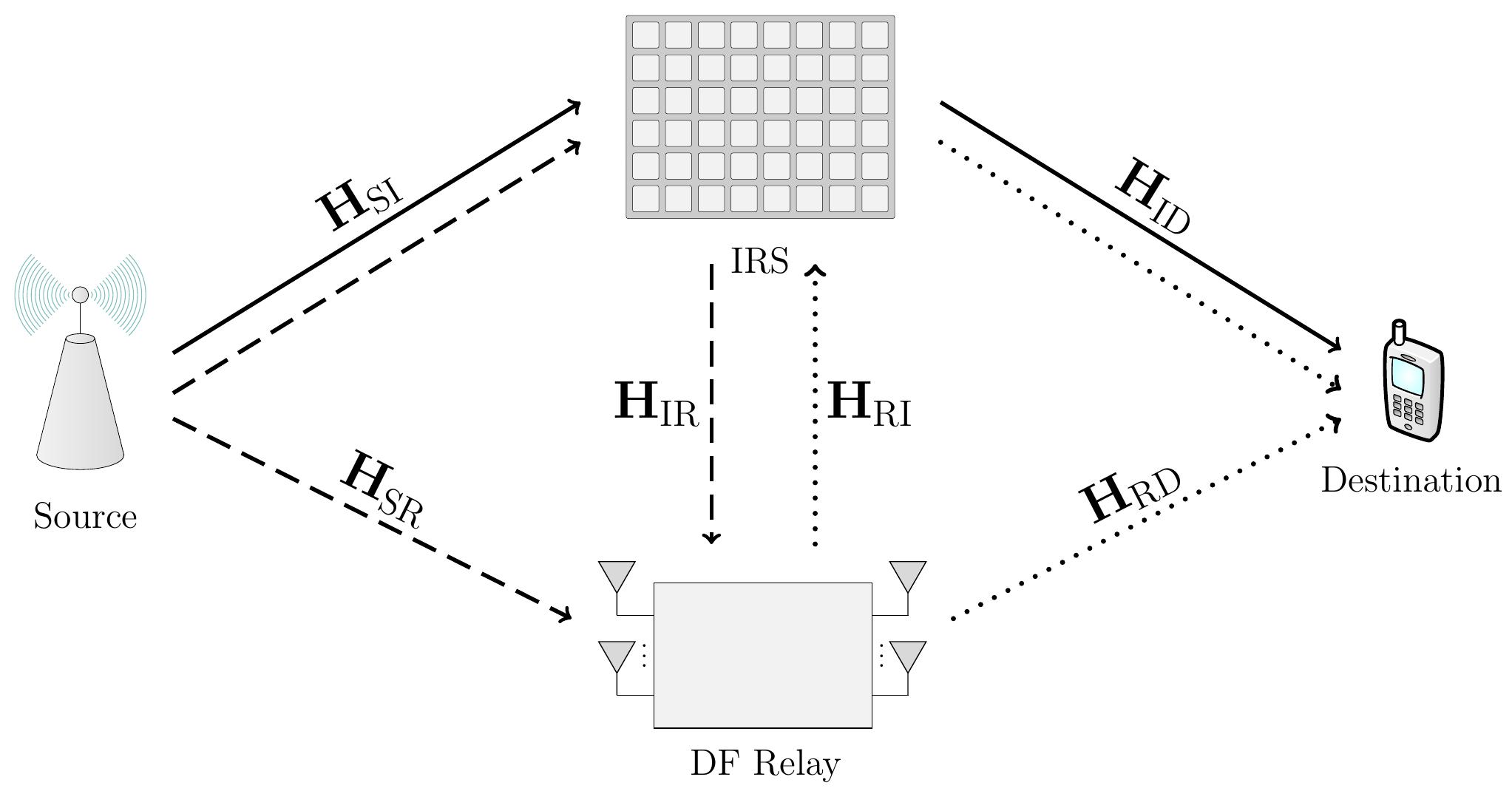}
  \caption{Two-stage \ac{irs}- and relay-assisted \ac{mimo} system. Solid arrows represent the S-D link in stage 1, dashed arrows the S-R link in stage 1, and dotted arrows the R-D link in stage 2.}
  \label{fig:system_model}
\end{figure}

We consider the narrowband single-user \ac{mimo} communication system shown in Fig.~\ref{fig:system_model}, wherein the transmission from a source (S) to a destination (D) is assisted by both a  relay (R) and an \ac{irs} (I). We assume that S and R have maximum transmit powers $P_{\rm S}$ and $P_{\rm R}$, respectively. 

S, D, and R are equipped with \acp{ula} with $N_{\rm S}$, $N_{\rm D}$, and $N_{\rm R}$ antennas, respectively, whereas I is a \ac{upa} with $N_{\rm I}$ passive reflective elements. 

We denote with $\bm{H}_{\rm SI} \in \mathbb{C}^{N_{\rm I} \times N_{\rm S}}$ and  $\bm{H}_{\rm SR} \in \mathbb{C}^{N_{\rm R} \times N_{\rm S}}$ the S-R and S-I channels, with $\bm{H}_{\rm RI} \in \mathbb{C}^{N_{\rm I} \times N_{\rm R}}$ and $\bm{H}_{\rm RD} \in \mathbb{C}^{N_{\rm D} \times N_{\rm R}}$ the R-I and R-D channels, and with $\bm{H}_{\rm IR} = \bm{H}_{\rm RI}^T$ and $\bm{H}_{\rm ID} \in \mathbb{C}^{N_{\rm D} \times N_{\rm I}}$ the I-R and I-D channels. We consider narrowband mmWave channels \cite{Rappaport2014}, each having $M$ \ac{nlos} components. 
Hence, channel matrix $\bm{H}_{{\rm XY}}$ between transmitter $\rm{X}$ and receiver $\rm{Y}$ is 
\begin{equation}\label{channelmatrix}
    \bm{H}_{\rm{XY}} = \frac{1}{\sqrt{M}}\sum_{m=1}^{M} g_m\rho(d)\bm{a}\left(\bm{\omega}_{{\rm X},m}\right)\bm{a}^H\left(\bm{\omega}_{{\rm Y},m}\right),
\end{equation}
where $g_m \sim \mathcal{CN}(0, 1)$ is the gain of the $m$-th path, $\rho(d)$ is the path loss attenuation factor, with $d$ being the distance between X and Y, $\bm{\omega}_{\cdot,m}=\left(\xi_{\cdot,m}, \psi_{\cdot,m}\right)$ is the vector of azimuth ($\xi_{\cdot,m}$) and elevation ($\psi_{\cdot,m}$) angles, and $\bm{a}\left(\bm{\omega}_{\cdot,m}\right)=\left(1,\ldots,e^{j\pi[x \sin(\psi_{\cdot,m})\cos(\xi_{\cdot,m})+y \sin(\psi_{\cdot,m})\sin(\xi_{\cdot,m})]},\ldots \right)^{T}$ is the array response vector for the $m$-th path, with $1 \leq x \leq N_{\rm{X}}-1$ and $1 \leq y \leq N_{\rm{Y}}-1$. We assume all devices operate in \ac{hd} and have perfect channel state information.  

\subsection{\ac{irs} Model}

Each element of the \ac{irs} acts as an omnidirectional antenna element that captures and reflects signals, introducing an attenuation and a phase shift on the baseband-equivalent signal. Following the model of \cite{Abeywickrama20}, we denote with $\phi_n=A_n(\theta_n)e^{j\theta_n}$ the reflection coefficient of the $n$-th \ac{irs} element, where $\theta_n \in [-\pi,\pi)$ is the induced phase shift and $A_n^2(\theta_n) \in [0,1]$ is the corresponding power attenuation factor. Indicating with $\bm{x} \in {\mathbb C}^{1\times N_{\rm I}}$ the impinging signal on the \ac{irs}, the reflected signal $\bm{y} \in {\mathbb C}^{1\times N_{\rm I}}$ is $\bm{y} = \bm{\Phi}\bm{x}$, with  $\bm{\Phi}=\diag(\phi_1,\ldots,\phi_{N_{\rm I}})$, which is the \ac{irs} reflection matrix, also denoted  {\em \ac{irs} configuration}.

We consider the realistic baseband-equivalent model of the \ac{irs} described in  \cite{Abeywickrama20}, where 
\begin{equation}
    A_n(\theta_n) = (1-A_{\rm \min})\left(\frac{\sin{\theta_n-\zeta}+1}{2}\right)^\nu + A_{\rm \min},
\end{equation}
with $A_{\rm \min} \geq 0$, $\zeta \geq 0$, and $\nu \geq 0$ being \ac{irs}-specific parameters, assumed to be identical for all \ac{irs} elements. The phase shifts $\theta_n$ are controllable, thus indirectly controlling also the attenuations. Moreover, since continuous-phase shifts are hardly implementable \cite{Tan2016}-\cite{Tan2018}, we assume that the phase shifts are chosen from a discrete set $\mathcal{F}_\theta = \left\{0, \frac{2\pi}{2^b},\ldots,\frac{2\pi(2^b-1)}{2^b} \right\}$, where $b>0$ is the \ac{irs} phase shift resolution, i.e., the number of bits employed to control the phase shifts. The source has full control of the phase shifts, which can be optimized together with beamforming.

\subsection{Two-stage Communication Protocol}

For a \ac{hd} \ac{df} relay, signal reception and transmission have to occur in two stages, here assumed to be of the same duration.
 
    \paragraph*{Stage 1}
    S splits the message into two sub-messages, and encodes/modulates them into the  two signals $\bm{x}_{\rm SR}$ and $\bm{x}_{\rm SD}$, intended for R and D, respectively. The two signals are precoded with  \ac{bd} precoders  $\bm{B}_{\rm SR}$ and $\bm{B}_{\rm SD}$ before transmission, such that they are received only at the indented destination, without mutual interference. The signal transmitted by S is thus
    \begin{equation}
        \bm{s} = \bm{B}_{\rm SR} \bm{x}_{\rm SR} + \bm{B}_{\rm SD} \bm{x}_{\rm SD}.
    \end{equation}
    Then, for a given IRS configuration $\bm{\Phi}_1$, the received signals at R and D are, respectively,
    \begin{equation}  \label{yR1} 
        \bm{y}_{R,1} = \underbrace{(\bm{H}_{\rm SR}+ \bm{H}_{\rm IR}\bm{\Phi}_1\bm{H}_{\rm SI})\bm{B}_{\rm SR}}_{\Tilde{\bm{H}}_{\rm SR}(\bm{\Phi}_1)} \bm{x}_{\rm SR} + \bm{n}_{R,1},
    \end{equation}
    \begin{equation}  \label{yD1} 
        \bm{y}_{D,1} = \underbrace{\bm{H}_{\rm ID}\bm{\Phi}_1\bm{H}_{\rm SI}\bm{B}_{\rm SD} }_{\Tilde{\bm{H}}_{\rm SD}(\bm{\Phi}_1)}\bm{x}_{\rm SD} + \bm{n}_{D,1},
    \end{equation}
    where $\Tilde{\bm{H}}_{\rm SR}(\bm{\Phi}_1)$ ($\Tilde{\bm{H}}_{\rm SD}(\bm{\Phi}_1)$) is the S-R (S-D) equivalent channel matrix (we highlight their dependency on the \ac{irs} configuration), and $\bm{n}_{R,1} \sim \mathcal{CN}\left(0,\sigma^2\bm{I}_{N_{\rm R}}\right)$ ($\bm{n}_{D,1} \sim \mathcal{CN}\left(0,\sigma^2\bm{I}_{N_{\rm D}}\right)$) is the complex Gaussian noise vector at R (D). 
    
    \paragraph*{Stage 2}
    S remains silent, while R decodes the sub-message received by S in stage 1 and re-encodes/re-modulates it into the signal $\bm{x}_{\rm RD}$. Then, R transmits $\bm{x}_{\rm RD}$ to D with the \ac{irs} using a new  configuration $\bm{\Phi}_2$. D receives the signal vector
    \begin{equation} \label{yD2}
        \bm{y}_{D,2} = \underbrace{(\bm{H}_{\rm RD}+\bm{H}_{\rm ID}\bm{\Phi}_2\bm{H}_{\rm RI})}_{\Tilde{\bm{H}}_{\rm RD}(\bm{\Phi}_2)}\bm{x}_{\rm RD} + \bm{n}_{D,2}, 
    \end{equation}
    where $\Tilde{\bm{H}}_{\rm RD}(\bm{\Phi}_2)$ is the R-D equivalent channel matrix, and $\bm{n}_{D,2} \sim \mathcal{CN}\left(0,\sigma^2\bm{I}_{N_{\rm D}}\right)$ is the complex Gaussian noise vectors at $D$.
    
Note that, in both stages, the \ac{irs} configurations $\bm{\Phi}_1$ and $\bm{\Phi}_2$ are provided by S, which has full control of the phase shifts.

\section{Maximum-Rate Problem}

We now first derive the \ac{ar} and then, we formulate the problem of maximizing the \ac{ar}.

\subsection{Achievable Rate}

For the first stage, the transmit beamformers $\bm{B}_{\rm SD}$ and $\bm{B}_{\rm SR}$ are chosen such that $\bm{x}_{\rm SR}$ and $\bm{x}_{\rm SD}$ do not generate interference at D and R, respectively. To this end, \ac{bd} is applied (see \cite{Tomasin}), using in general a reduced set of streams for the two links. Let $\bm{H}_{\rm SD} = \bm{U}_{\rm SD}\bm{\Gamma}_{\rm SD} \bm{V}_{\rm SD}$ and the \ac{svd} of $\bm{H}_{\rm SD}$; a subset $\mathcal S_{\rm SD}$ of streams (corresponding to diagonal elements of $\bm{\Gamma}_{SD}$) is selected for transmission to D. The  \ac{bd} beamformer for transmission to R is $\bm{B}_{\rm SR} = \bm{N}_{\rm SD}\bm{B}'_{\rm SR}$, where  $\bm{N}_{\rm SD}$ collects the columns of $\bm{V}_{\rm SD}$ with indices not in the set $\mathcal S_{\rm SD}$, while $\bm{B}'_{\rm SR}$ is  the capacity-achieving precoder for the resulting S-R channel. A similar procedure is applied for the definition of the S-D precoder $\bm{B}_{\rm SD}$, for which $\mathcal S_{\rm SR}$ streams are selected. We must also have $|\mathcal S_{\rm SR}| + |\mathcal S_{\rm SR}| \leq N_{\rm S}$. Lastly,  $\bm{x}_{SD}$ and $\bm{x}_{SR}$, are zero-mean complex Gaussian vectors with independent entries of size  $|\mathcal S_{\rm SD}|$ and $|\mathcal S_{\rm SR}|$. 

For the second stage, R applies capacity-achieving precoding, and  $\bm{x}_{RD}$ zero-mean complex Gaussian vectors with independent entries of size $N_{\rm R}$.

As a result, the S-D \ac{mimo} equivalent channel can be decomposed into $|\mathcal{S}_{\rm SD}|$ independent parallel \ac{awgn} channels with gains $\{\gamma_{\rm SD}(i)\}$.\footnote{$\gamma_{\rm SD}(i)$ is the $i$-th singular value of $\Tilde{\bm{H}}_{\rm SD}(\bm{\Phi}_1)\Tilde{\bm{H}}_{\rm SD}^H(\bm{\Phi}_1)$.}  The capacity of the S-D channel is therefore 
\begin{equation}
    C_{\rm SD} = \sum_{i \in \mathcal{S}_{\rm{SD}}} \log_2\left[1+\gamma_{\rm SD}(i) \frac{P_{\rm SD}(i)}{\sigma^2}\right],
    \label{capa}
\end{equation}
where $P_{\rm SD}(i)$ is the power allocated to channel $i$. Similarly, the S-R and R-D channels can be decomposed into $|\mathcal{S}_{\rm SR}|$ and $|\mathcal{S}_{\rm RD}|$ parallel \ac{awgn} channels, with gains $\{\gamma_{\rm SR}(i)\}$ and $\{\gamma_{\rm RD}(i)\}$, respectively, and   the S-R and R-D capacities $C_{\rm SR}$ and $C_{\rm RD}$ can be written as in \eqref{capa}, where subscript SD is replaced by subscripts SR and RD, respectively.

The \ac{ar} of the considered two-stage scheme is therefore
\begin{equation}
    C_{\rm HYB} = \frac{1}{2} (C_{\rm SD} + \min\{C_{\rm SR}, C_{\rm RD}\}),
    \label{CRSR}
\end{equation}
where the two stages  requires twice the time of direct transmission, hence the factor 1/2.

Note that for a transmission using only the relay, the \ac{ar}
$C_{\rm relay}$ is still given by \eqref{CRSR}, with the \ac{irs} switched off ($A_n(\theta) = 0$, $\forall \theta$). A transmission using only the \ac{irs} can instead be performed in a single stage and the \ac{ar} is $C_{\rm \ac{irs}} =  C_{\rm SD}$. In both cases, no \ac{bd} is needed.
Note also that IRS- or relay-only transmissions  occur if no streams are selected for the S-R or S-D links, i.e., if $|\mathcal{S}_{\rm SR}|=0$ or $|\mathcal{S}_{\rm SD}|=0$, respectively.

\subsection{Optimization Problem}

With this choice of beamformers, we are left with the problem of optimizing a) the transmit power, b) the \ac{irs} configurations in both stages, and c) the set of streams assigned to R and D in stage 1. The \ac{ar} maximization problem can be formalized as follows:
\begin{subequations}
    \label{maxprobl}
    \begin{equation}
            \argmax_{\substack{\bm{\Phi}_1, \bm{\Phi}_2 \\\mathcal{S}_{\rm SD}, \mathcal{S}_{\rm SR}, \mathcal{S}_{\rm RD}\\\left\{P_{\rm SD}(i)\right\},\left\{P_{\rm SR}(j)\right\},\left\{P_{\rm RD}(k)\right\}}} \hspace{-1cm} \left( C_{\rm SD} + C_{\rm SR} \right), \label{target}
    \end{equation}
    ~\vspace{-0.4cm}
    \begin{alignat}{2}
    & \text{s.t.}\; &      & \bm{\Phi}_k=\diag(\phi_{1,k},\ldots,\phi_{N_{\rm I},k}), \quad k=1,2, \label{Cphi1}\\
    &                   &      & \phi_{n,k}=A_{n,k}(\theta_{n,k})e^{j\theta_{n,k}}, \quad 1 \leq n \leq N_{\rm I}, \quad k=1,2, \label{con_phi}\\
    &                   &      & \theta_{n,k} \in \mathcal{F}_\theta, \quad 1 \leq n \leq N_{\rm I}, \quad k=1,2, \label{Cth2}\\
    &                   &      & \sum_{i \in \mathcal{S}_{\rm{SD}}} P_{\rm SD}(i) + \sum_{j \in \mathcal{S}_{\rm{SR}}} P_{\rm SR}(j) \leq P_{\rm S}, \label{CP0}\\
    &                   &      &   \sum_{k\in \mathcal{S}_{\rm RD}} P_{\rm RD}(k) \leq P_{\rm R}, \label{CP2}\\
    &                   &      & \sum_{i \in \mathcal{S}_{\rm{SD}}} P_{\rm SD}(i) + \sum_{j \in \mathcal{S}_{\rm{SR}}}P_{\rm SR}(j) + \sum_{k \in \mathcal{S}_{\rm{RD}}} P_{\rm RD}(k)  \leq P_{\rm max}, \label{CP1}\\
    &                   &      & C_{\rm SR} \leq C_{\rm RD}  \label{consSRRD} \\
    & & &\mathcal S_{\rm SD}, \mathcal S_{\rm SR}  \in \{1, \ldots, N_{\rm S}\}, \label{constre1}\\
        & & & 0<|\mathcal S_{\rm SD}| + |\mathcal S_{\rm SR}| \leq N_{\rm S}. \label{constre2} 
    \end{alignat}
\end{subequations}
The minimum in \eqref{CRSR} is now reflected by constraint \eqref{consSRRD}. Constraints \eqref{Cphi1}-\eqref{Cth2}  are related to the control of \ac{irs} phase shifts, and constraints \eqref{CP0} and \eqref{CP1} are power constraints at the devices, and we added the total power constraint $P_{\max}$. This constraint will make the comparison with schemes using only the \ac{irs} more fair, by imposing $P_{\max}$ the maximum power for S.  Constraints \eqref{constre1}-\eqref{constre2} are relative to the stream assignment.
 
\subsection{Alternating Optimization Solution}
Notice that constraint \eqref{con_phi} makes the problem non-convex, thus we resort to an alternating optimization solution, where we optimize over the \ac{irs} configurations and stream sets, and for each considered configuration we optimize the transmission powers.

For fixed \ac{irs} configurations and stream selections, the optimization problem \eqref{maxprobl} becomes
\begin{equation}
         \argmax_{\substack{\left\{P_{\rm SD}(i)\right\},\left\{P_{\rm SR}(j)\right\},\left\{P_{\rm RD}(k)\right\}}} \hspace{-1cm}   \left( C_{\rm SD} + C_{\rm SR} \right),  \quad 
    \mbox{s.t. } \eqref{CP0}-\eqref{consSRRD}.
\label{optProb}
\end{equation}
Observe that, due to constraint \eqref{consSRRD}, the problem is still non-convex. However, the powers in stage 1 and stage 2 are coupled only through the constraint \eqref{CP1}. We can decouple the two problems by introducing the auxiliary variable $P_{\rm R, eff}$ such that 
\begin{equation}
  \sum_{k \in \mathcal{S}_{\rm{RD}}} P_{\rm RD}(k) =  P_{\rm R, eff},
\end{equation}
so that the power that can be effectively used by S is, from  \eqref{CP0}, \eqref{CP2}, and \eqref{CP1},  as $
    P_{\rm S, eff} = \min\{P_{\rm S}, P_{\rm max} - P_{\rm R, eff}\}$.
With these new definitions, \eqref{optProb} can be split into the two (coupled) problems, for given $P_{\rm R, eff}$, 
\begin{subequations}
\label{P1}
    \begin{alignat}{2}
    C_{\rm RD}^*=  & \max_{\left\{P_{\rm RD}(k)\right\}} & \, & C_{\rm RD},  
    \quad
    \text{s.t.}   \sum_{k \in \mathcal{S}_{\rm{RD}}} P_{\rm RD}(k) = P_{\rm R, eff},
    \end{alignat}
\end{subequations}
and
\begin{subequations}\label{opt1}
\begin{equation}
    \argmax_{\substack{\left\{P_{\rm SD}(i)\right\},\left\{P_{\rm SR}(j)\right\}}} 
    \frac{1}{2}\left( C_{\rm SD} + C_{\rm SR} \right), 
\end{equation}  
    \begin{alignat}{2}
    & \text{s.t.} \;&      & C_{\rm SR} \leq C_{\rm RD}^*, \label{capConstP2} \\
    &                   &      & \sum_{i \in \mathcal{S}_{\rm{SD}}} P_{\rm SD}(i) + \sum_{j \in \mathcal{S}_{\rm{SR}}} P_{\rm SR}(j) = P_{\rm S, eff}. \label{PSeffconst}
    \end{alignat}
    \label{probopt1}
\end{subequations} 
Note that \eqref{P1} and \eqref{opt1} are convex optimization problems and, therefore, they can be solved in closed-form, as detailed in the next sub-section.

Then, we need to optimize  the \ac{irs} reflection coefficients $\bm{\Phi}_1$ and $\bm{\Phi}_2$, the stream sets $\mathcal{S}_{\rm SR}$, $\mathcal{S}_{\rm SD}$, and $\mathcal{S}_{\rm RD}$, and the auxiliary variable $P_{\rm R, eff}$, in what turns out to be a non-convex problem. Thus, we resort to the discrete \ac{ga} \cite{Holland1972}, which operates iteratively, solving  sub-problems \eqref{opt1} and \eqref{P1} for given \ac{irs} configurations, power $P_{\rm R, eff}$, stream sets $\mathcal{S}_{\rm SR}$, $\mathcal{S}_{\rm SD}$, and $\mathcal{S}_{\rm RD}$ at each iteration.

\subsection{Decoupled Problem Solution}

\paragraph*{Solution of Problem \eqref{P1}}

Since the capacity $C_{\rm SR}$ is upper bounded by $C_{\rm RD}^*$ from \eqref{capConstP2}, we first  optimize the transmit powers $\left\{P_{\rm RD}(k)\right\}$ at R, given  $P_{\rm R, eff}$. Indeed, \eqref{P1} can be solved via the  standard waterfilling algorithm \cite{Tomasin} on channels with gains $\{\gamma_{\rm RD}(i)\}$ and total power $P_{\rm R, eff}$.
 
\paragraph*{Solution of Problem \eqref{opt1}} 
The Lagrangian function of \eqref{opt1} is (with $\lambda_1$ and $\lambda_2$ multipliers)
\begin{align}
    \mathcal{L} & =  \left( C_{\rm SD} + C_{\rm SR} \right)\notag -\lambda_2 \left(C_{\rm SR} -C_{\rm RD}^* + s\right) \\
    & - \lambda_1 \left(\sum_{i \in \mathcal{S}_{\rm{SD}}} P_{\rm SD}(i) + \sum_{j \in \mathcal{S}_{\rm{SR}}} P_{\rm SR}(j) - P_{\rm S, eff}\right),
\end{align}
where $s \geq 0$ is an additional slack variable. Setting to zero the derivative of the Lagrangian function, we obtain  the following stationary points
\begin{align}
     P_{\rm SD}(i) = \frac{1}{\ln(2) \lambda_1} - \frac{1}{\gamma_{\rm SD}(i)}, \; 
     P_{\rm SR}(j) = \frac{1}{\ln(2) \lambda_1} - \frac{1}{\gamma_{\rm SR}(j)},
\end{align}
with $\lambda_1$ such that \eqref{PSeffconst} is satisfied.

Now, letting $s = C_{\rm RD}^* - C_{\rm SR}$, if $s \geq 0$ we have found the optimal solution. If instead $s < 0$, then we must assume $s=0$, i.e., the S-R rate in stage $1$ equals the R-D rate in stage $2$. Consequently, we allocate the minimum power that satisfies this constraint to the S-R link, while all the remaining power is assigned to the S-D link. Hence, we first solve the following problem
\begin{equation}
    \argmin_{\left\{ P_{\rm SR}(j)\right\}}  \sum_{j \in \mathcal{S}_{\rm{SR}}} P_{\rm SR}(j),  \quad \mbox{\, s.t.\, }  C_{\rm SR} =C_{\rm RD}^*,  
    \label{subprob1}
\end{equation}     
with the Lagrangian multipliers method, providing 
\begin{equation}
    P^*_{\rm SR}(j) = \left[\left(\frac{2^{C_{\rm RD}^*}}{\prod_{j\in \mathcal{S}_{\rm SR}} \gamma_{\rm SR}(j)}\right)^{\frac{1}{|\mathcal{S}_{\rm SR}|}} - \frac{1}{\gamma_{\rm SR}(j)}\right]^+,
\end{equation}
where $(x)^+ = x$ if $x \geq 0$, while $(x)^+ = 0$ otherwise.
For the obtained optimal powers $P_{\rm SR}(j)^*$, we solve 
\begin{equation}
     \argmax_{\left\{P_{\rm SD}(i)\right\}}  \,  C_{\rm SD}, \quad  \mbox{\, s.t.\, } \eqref{PSeffconst},
    \label{subprob2}
\end{equation}
which  is similar to \eqref{P1} and can be solved likewise.

\section{Numerical Results}
In this section, we assess the performance of the proposed protocol.  S, R, D, and I have coordinates  $(0,0,3)$, $(10,-10,3)$, $(20,0,1.5)$, and $(10,y_{\rm I},3)$~m, respectively (see Fig.~\ref{fig:system_model}), and $y_{\rm I}$ is a parameter to be set. 
We consider $M=2$ \ac{nlos} components for each mmWave link. S, R, and D are equipped with \acp{ula} of $N_{\rm S}=16$, $N_{\rm R}=8$, and $N_{\rm D}=4$ antennas, respectively, whereas the \ac{irs} is an \ac{upa} with $N_{\rm I}=36$ elements and parameters (see \cite{Abeywickrama20}) $A_{\rm min}=0.2$, $\zeta=0.43 \pi$, and $\nu=1.6$.  Angles in the array response vector are chosen according to a uniform random distribution, in particular, $\psi_{\cdot, m} \sim \mathcal{U}[0, 2\pi)$ and $\xi_{{\rm I},m} \sim \mathcal{U}[0, \pi/2)$ for the \ac{irs}, while $\xi_{\cdot, m} = 0$ for other devices with \ac{ula}.
The transmit \ac{snr} is $P_{\rm max}/\sigma^2 = 10$ ($10$ dB). The path loss term is modelled as $\rho(d) = K_0 (d/d_0)^{-\alpha/2}$, where $K_0 = \rho(d_0) = 0$ dB is the path loss at the reference distance $d_0=10$~m, and $\alpha=5.76$ is the path loss exponent \cite{Rappaport2013}. We compare five schemes: the proposed optimized hybrid \ac{irs}-relay scheme ({\tt Hyb. Opt.}), a hybrid scheme with random \ac{irs} configuration ({\tt Hyb. Rand.}), a scheme without relay and an optimized \ac{irs} ({\tt \ac{irs} Opt.}), a scheme with a random \ac{irs} ({\tt \ac{irs} Rand.}), and  a scheme without \ac{irs} and a relay ({\tt Relay}).

\begin{figure}
    \centering
    \includegraphics[width=0.95\columnwidth]{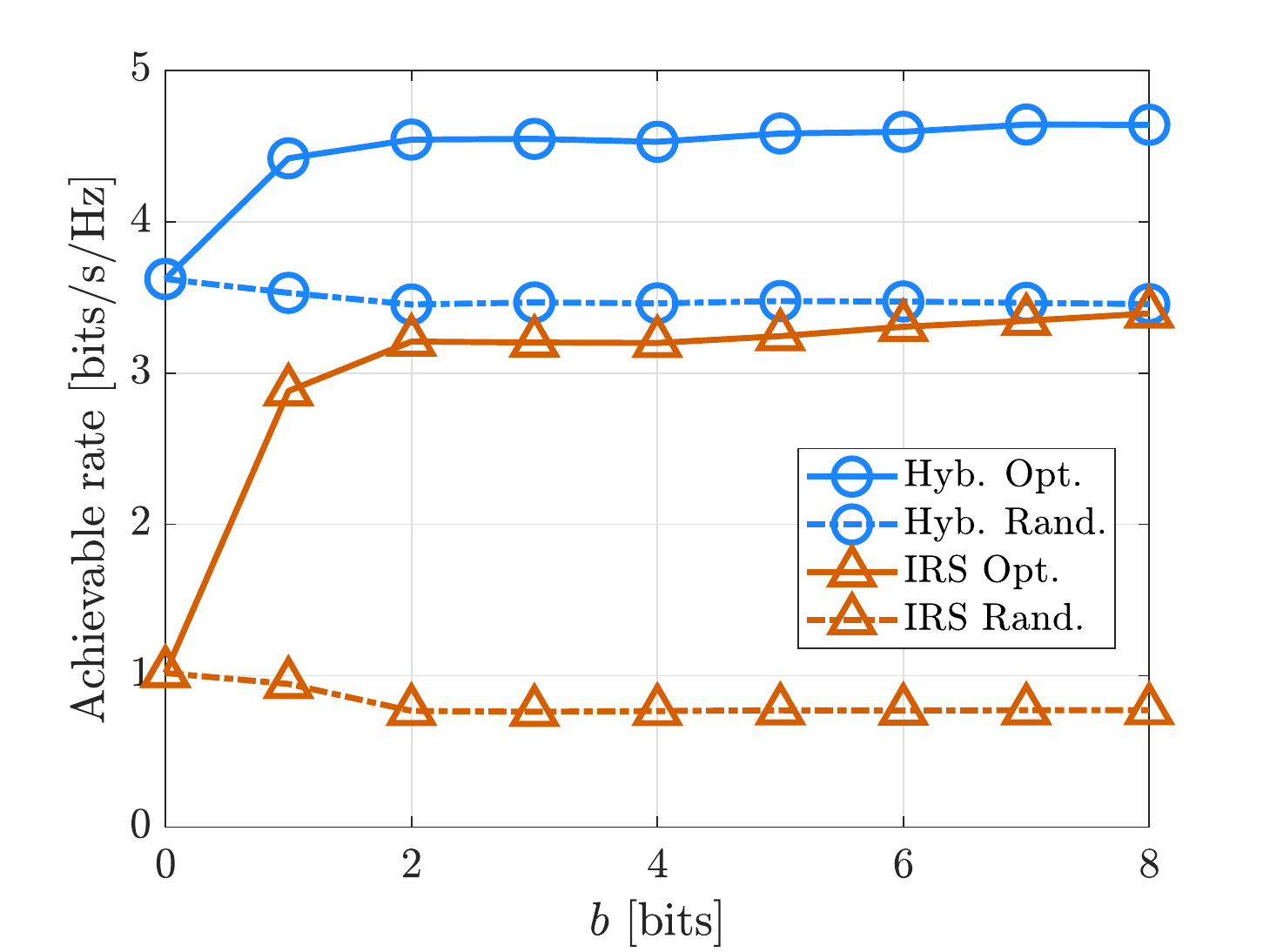}
    \caption{\ac{ar} versus $b$, for $P_{\rm S}/P_{\rm max} = 0.5$, $N_I=36$, and $y_{\rm I}=20$~m.}
    \label{fig:f_of_b}
\end{figure}

Fig.~\ref{fig:f_of_b} shows the \ac{ar} as a function of the \ac{irs} phase shift resolution $b$ for $y_{\rm I}=20$~m and  $P_{\rm S}/P_{\rm max} = 0.5$. For $b=0$ we consider a fixed \ac{irs} configuration with phase shifts $\theta_n = \pi$, $\forall n$, corresponding to the maximum value of $A(\cdot)$. For all schemes, the \ac{ar} saturates with just $b=1$ or $2$ bits per element, thus, as already observed in the literature, a very limited number of configurations are enough to achieve the gains provided by the \ac{irs}. In the following, we will consider $b=2$. The schemes with randomly configured \ac{irs} show a penalty for higher resolution, since configurations with lower gains $A(\cdot)$ are used.

\begin{figure}
    \centering
    \includegraphics[width=0.95\columnwidth]{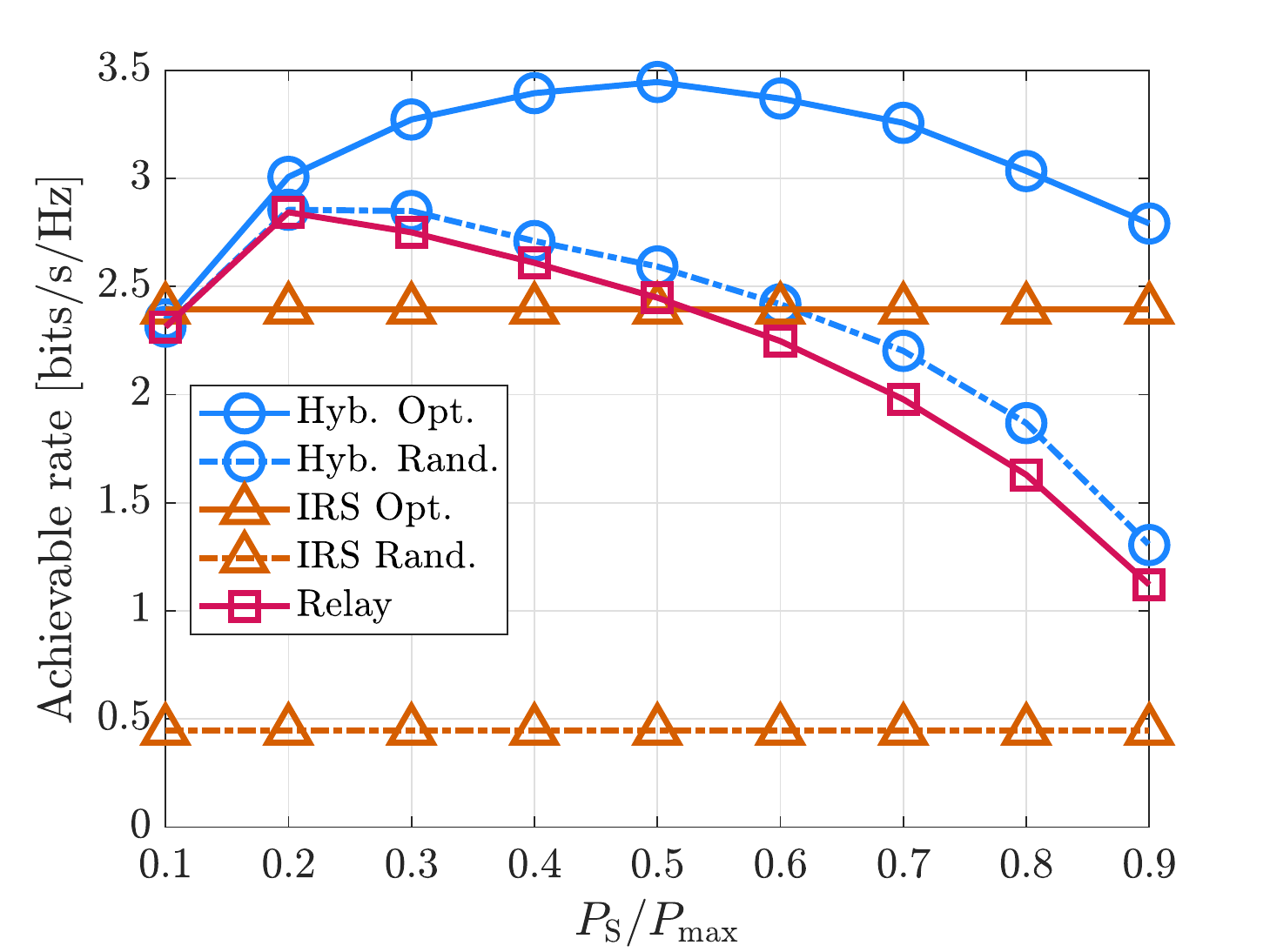}
    \caption{\ac{ar} versus $P_{\rm S}/P_{\rm max}$, for $y_{\rm I}=20$~m, $N_I=36$, and $b=2$.}
    \label{fig:f_of_P_S}
\end{figure}

Fig.~\ref{fig:f_of_P_S} shows the \ac{ar} as a function of the fractional  available power at S, i.e., $P_{\rm S}/P_{\rm max}$ for $y_{\rm I}=20$~m. The {\tt Hyb. Opt.} scheme provides the highest \ac{ar} for all values of $P_{\rm S}/P_{\rm max}$. Still, for low  $P_{\rm S}/P_{\rm max}$, the relay has a considerable fraction of power, thus the {\tt Relay} scheme is close to optimal. Instead, at high $P_{\rm S}/P_{\rm max}$, the constraint on $C_{\rm RD}$ limits the \ac{ar} at the relay, and the {\tt \ac{irs} Opt.} scheme attains higher performance. The {\tt \ac{irs} Rand.} scheme yields very poor performance, due to the absence of the relay and the random configuration of the \ac{irs}.
\begin{figure}
    \centering
    \includegraphics[width=0.95\columnwidth]{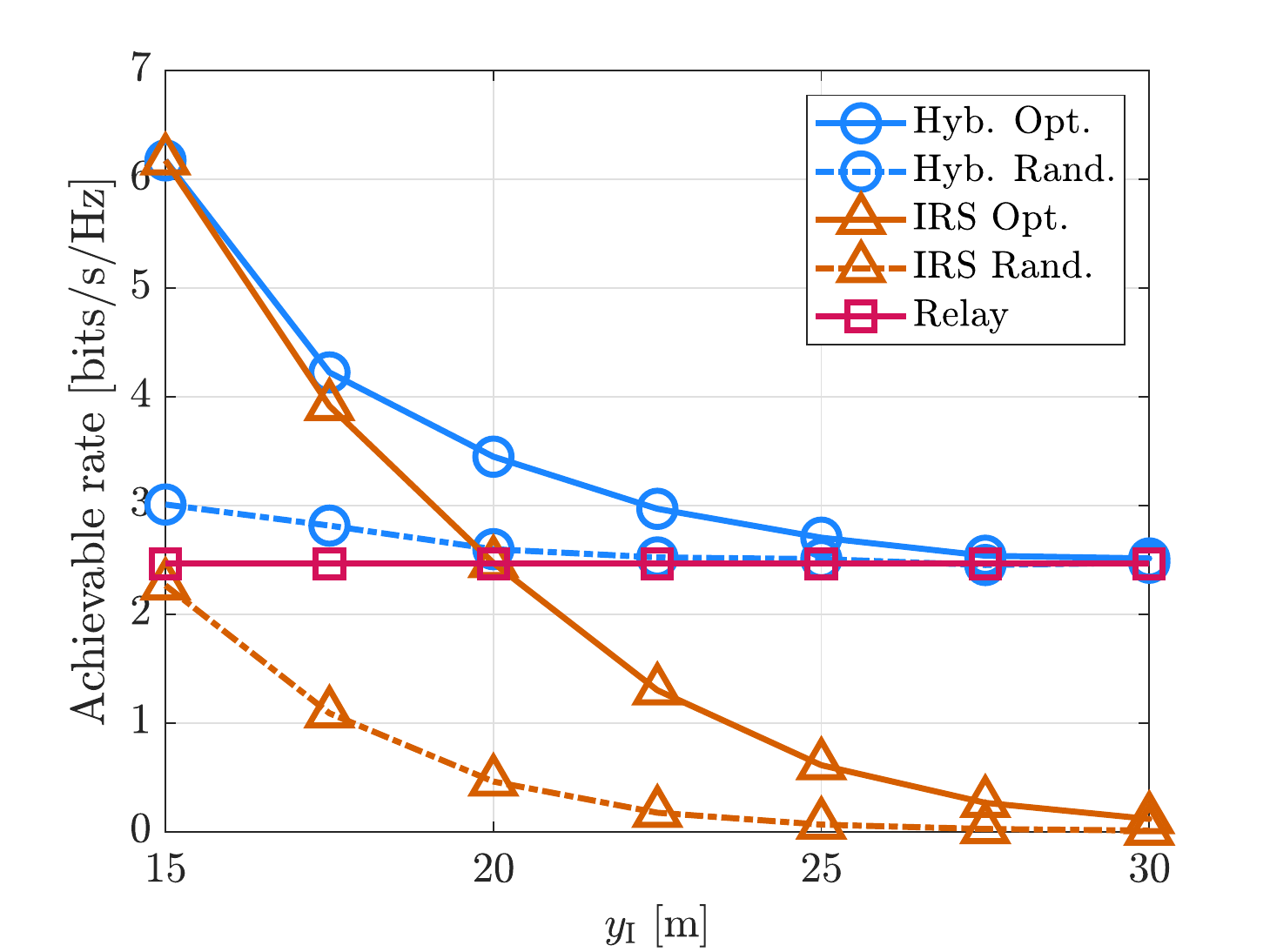}
    \caption{\ac{ar} versus $y_{\rm I}$, for $P_{\rm S}/P_{\rm max} = 0.5$, $N_I=36$, and $b=2$.}
    \label{fig:f_of_y_I}
\end{figure}

Fig.~\ref{fig:f_of_y_I} shows the \ac{ar} as a function of the \ac{irs} distance $y_{\rm I}$, when $P_{\rm S}/P_{\rm max} = 0.5$. 
For small $y_{\rm I}$ values, the \ac{irs} link is dominant with respect to the relay link, making the {\tt Hyb. Opt.} scheme transmit exclusively towards the \ac{irs}, thus avoiding the half-rate penalty of the two-stage protocol, and approaching the \ac{ar}. On the other hand, the \ac{irs} assistance becomes marginal as $y_{\rm I}$ grows, resulting in similar performance between {\tt Hyb.} and {\tt Relay} schemes.
 
\begin{figure}
    \centering
    \includegraphics[width=0.95\columnwidth]{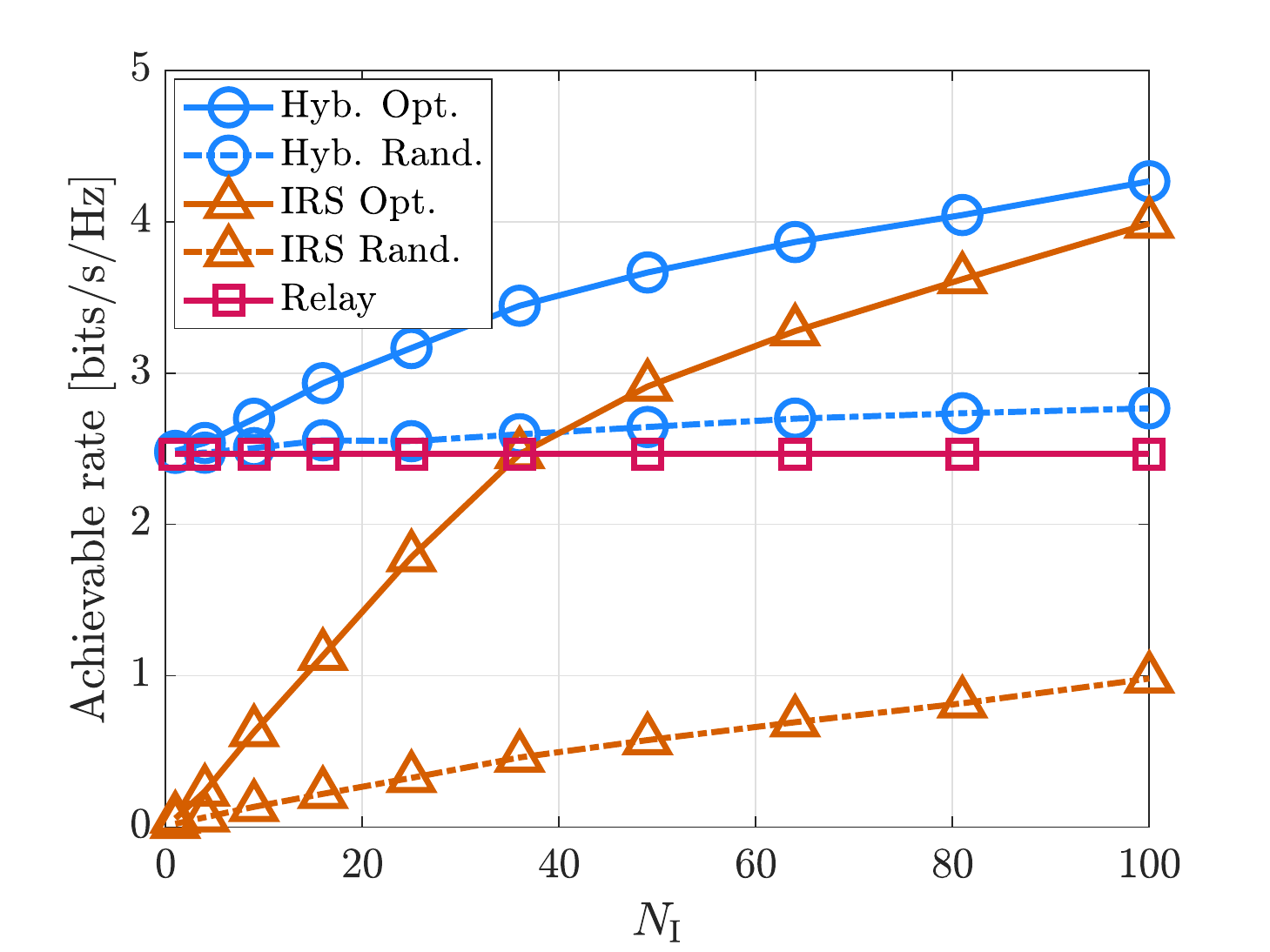}
    \caption{\ac{ar} versus $N_{\rm I}$, for $P_{\rm S}/P_{\rm max} = 0.5$, $y_{\rm I} = 20$~m and $b=2$.}
    \label{fig:f_of_N_I}
\end{figure}

Finally, Fig.~\ref{fig:f_of_N_I} shows the \ac{ar} as a function of the number of reflecting elements $N_{\rm I}$, for $P_{\rm S}/P_{\rm max} = 0.5$ and $y_{\rm I} = 20$~m. As expected, due to the huge beamforming gain introduced by large \acp{irs}, the \ac{ar} grows with the number of reflecting elements.

\section{Conclusions}

In this paper, we considered an hybrid \ac{irs}-relay system, optimizing  power allocation, \ac{irs} configurations, and stream sets to maximize the \ac{ar}. Numerical results showed that, in the considered scenarios, large phase-optimized \acp{irs} yield higher \acp{ar} than systems using only either the relay or the \ac{irs}. 
Indeed, the best performance is achieved by different uses of the relay and the \ac{irs} under different positions of the devices or power split among the source and the relay.
This suggests that the proposed hybrid solution, which is able to switch among the various uses, is always advantageous.

\bibliographystyle{IEEEtran}
\bibliography{Bibliography.bib}
\end{document}